\documentstyle[12pt]{article}
\begin{document}

\title{\Large \bf Deformed Poincar\'e Algebra and Field Theory
\thanks{Partially supported by CEC Science project No. SC1-CT91-0729}}
\author{ {\normalsize Alexandros A. Kehagias},
 {\normalsize Patrick A. A.
Meessen}
 \thanks
 {On leave from
   Inst. Theor. Phys.,  Univ. of Nijmegen, 6525 ED Nijmegen, The Netherlands }
\\ [.1cm]
{\normalsize and} \\ [.1cm]
{\normalsize George Zoupanos}
\\ {\normalsize Phys. Dept., National Technical Univ.}
\\ {\normalsize 15773 Zografou Athens, Greece}}
\date{}
\maketitle
\begin{abstract}
\begin{sloppypar}
\normalsize
We examine deformed Poincar\'e algebras containing the  exact Lorentz algebra.
We impose constraints which are necessary for defining field theories
on these algebras and we present simple field theoretical
examples. Of particular
interest is a case that exhibits improved renormalization properties.
\end{sloppypar}
\end{abstract}



\newpage

\section{Introduction}

Deformations of space-time and its symmetries have attracted a lot of attention
recently \cite{1}--\cite{12}. The main reason from the particle physics
perspective is that such
deformed spaces or symmetries could be the basis to construct field theories
with improved ultraviolet properties. This hope was motivated from the fact
that q-deformations of space-time seem to lead to some lattice pattern which
in turn could serve at least as some kind of regularization built in a theory
that would be  defined on such space-time.

Deformations of the Poincar\'e algebra (dPA) have been considered so
far along three directions. The first consists of direct q-deformations
of the Lorentz sub-algebra \cite{1}--\cite{3}
along the lines prescribed by Drinfeld and
Jimbo \cite{11}. The second is based on the fact that the Poincar\'e algebra
(PA)
can be obtained by a
Wigner-In\"{o}n\"{u} contraction of the simple anti-de Sitter algebra
$O(3,2)$ \cite{4,5}.
Then one first constructs the q-deformed $O_q(3,2)$ using the
Drinfeld-Jimbo method and then does the contraction.
 In fact it was shown in ref.\cite{8} that the same deformation
can be obtained directly by considering general deformations of the commutation
relations in the PA. Unfortunately the above deformations do not preserve the
Lorentz algebra.
Therefore it is natural to search for those dPAs that leave the Lorentz
algebra unchanged in order to facilitate the quantization of the corresponding
field theories. This motivation led us in ref.\cite{10} to consider
 a third direction, namely
deformations of the PA that leave the Lorentz algebra invariant.

In the present paper we continue our search for the appropriate dPAs that will
allow us to construct field theories with impoved ultraviolet properties.
We demand that the dPAs,  in addition to leaving  the Lorentz
algebra invariant and giving the ordinary PA in low energies,  should
satisfy two more constraints.
First, we require that there exists a tensor product of representations
(coproduct) which is necessary in order to be able to go from the irreducible
representations in the Hilbert space of quantum mechanics to the
reducible representations in the
Fock space of free quantum fields.
 A second requirement is that the
 representations of the dPA should be different from those of the
 ordinary PA, a usual property in q-groups \cite{13},
as a way to guarantee
that the dPAs are well distinguished  from ordinary PA and, in
 principle,  with  different
physical implications. Finally we demonstrate using a scalar field theory
defined on a specific dPA that indeed field theories with improved ultraviolet
properties can be constructed.

\section{The deformed Poincar\'e algebra}
In ref.\cite{10} it has been proposed to search for deformations
of the PA that do not affect the
Lorentz subalgebra.

The Lorentz algebra is a six-dimensional Lie algebra
generated by the generators $J_i,\, K_i$ of rotations
and boosts correspondingly satisfying the following commutation relations
\begin{eqnarray}
{}[J_i,J_j]&=&i \epsilon_{ijk}J_k,  \nonumber \\
{}[J_i,K_j]&=& i\epsilon_{ijk} K_k,  \nonumber \\
{}[K_i,K_j]&=& -i \epsilon_{ijk}J_k .
\end{eqnarray}
Recall that by defining
\[
N_i=\frac{1}{2}(J_i+iK_i)\,
\]
one finds that $N_i$'s and $N_i^{\dag}$'s satisfy an $SU(2)\otimes SU(2)$
algebra.
The enlargement of the Lorentz algebra to the Poincar\'e by  including
the energy-momentum generators $(P_0,P_i)$  was
proposed  as a way to describe the quantum states
of relativistic particles as unitary representations of the Poincare\'e
group without using the wave equations \cite{14}.
One of the main points of ref.\cite{10}was to show that
this enlargement of the Lorentz algebra is not unique.
Indeed, it was proposed to introduce a generalized
set of commutation relations (as compared to the ordinary PA)
for the generators $(P_0,P_i,K_i)$ as follows
\begin{eqnarray}
{}[K_i,P_0]&=&i \alpha_i(P_0,\vec{P}), \nonumber \\
{}[K_i,P_j]&=& i\beta_{ij}(P_0,\vec{P}) ,
\end{eqnarray}
where $\alpha_i,\beta_{ij}$ are functions of $P_0$,
$P_i$. Then applying the Jacobi identities on the sets
$(J_i,K_i,P_0)$ and $(J_i,K_j,P_k)$ it was found
that the general form of $\alpha_i$ and $\beta_{ij}$ is
\begin{eqnarray}
\alpha_i(P_0,\vec{P})&=& \alpha(P_0,\vec{P})P_i,  \nonumber \\
\beta_{ij}(P_0,\vec{P})&=&
\beta(P_0,\vec{P})\delta_{ij}+\gamma(P_0,\vec{P})P_iP_j
.\end{eqnarray}
Assuming furthermore that there exists a Casimir invariant of the enlarged
algebra of the form
\begin{equation}
f(P_0)-\vec{P}^2,
\end{equation}
it was found that
\begin{eqnarray}
\alpha_i(P_0,\vec{P})&=&\alpha(P_0)P_i, \nonumber \\
\beta_{ij}(P_0,\vec{P})&=& \beta(P_0)\delta_{ij} .
\end{eqnarray}
Moreover the closure of the enlarged algebra required that
\begin{equation}
\alpha(P_0)\beta^{\prime}(P_0)=1 . \label{1}
\end{equation}
 In this way a minimally deformed Poincar\'e algebra was constructed
with commutation relations
\begin{eqnarray}
{}[J_i,J_j]&=&i \epsilon_{ijk}J_k,  \nonumber \\
{} [J_i,K_j]&=& i\epsilon_{ijk} K_k,  \nonumber \\
{} [K_i,K_j]&=& -i \epsilon_{ijk}J_k ,\nonumber \\
{} [J_i,P_0]&=&0, \nonumber \\
{} [J_i,P_j]&=& i\epsilon_{ijk} P_k,  \nonumber \\
{} [P_0,P_i]&=&0,  \nonumber \\
{} [K_i,P_0]&=&i \alpha(P_0)P_i, \nonumber \\
{}[K_i,P_j]&=& i\beta(P_0)\delta_{ij} , \label{c}
\end{eqnarray}
where $\alpha(P_0),\beta(P_0)$ satisfy eq.(\ref{1}).
 Note that the ordinary PA is
 obtained when $\alpha(P_0)=1$ and $\beta(P_0)=P_0$.

The above dPA has two Casimir invariants. One corresponds to the length of the
Pauli-Lubanski four-vector
\begin{equation}
W^2=W_0^2-\vec{W}\cdot \vec{W} ,
\end{equation}
where
\begin{eqnarray}
W_0&=&\vec{J}\cdot \vec{P}, \nonumber \\
W_i&=&\beta(P_0)J_i+\epsilon_{ijk}P_jK_k ,
\end{eqnarray}
with eigenvalues
\[
W^2=-\mu^2 s(s+1),
\]
where $s=0,1/2,\ldots $ is the spin.

The other Casimir invariant of the dPA corresponds to the $(mass)^2$ of
 the ordinary PA and it is given by
\begin{equation}
\beta^2(P_0)-\vec{P}\cdot \vec{P} =\mu^2.
\end{equation}
Let us also recall that the transformation $P_0\rightarrow \beta(P_0)$
 reduces the dPA to the ordinary PA for the set of generators $(J_i,K_i,P_i,
\beta(P_0))$.

\section{Constraints on the dPA parameter functions}
{}From the construction of the dPA discussed above, it is clear that the
functions $\beta(P_0)$ and,
 consequently, $\alpha(P_0)$  are not specified.
An obvious physical requirement that these functions should satisfy
 is to  let us obtain the ordinary
PA as a limit of the  dPA in low energies.
 Therefore we require that the low energy behaviour
 of $\beta(P_0)$ should be
\[
\beta(P_0)\sim P_0
\]
and therefore
\[
\alpha(P_0)\sim 1 .
\]

One of our main aims is to define  field theories  on the
constructed dPA hopefully with improved ultraviolet
 properties. A dPA with the
Lorentz invariant subalgebra paves the way for an easy first quantization
of such theories. Another  requirement for construction of field theories
is to be able to define the multiparticle states. In field theories defined
on the ordinary PA the multiparticle states are constructed from the tensor
product of one particle states. Here therefore we are looking for the
 corresponding "tensor" product which is usually called coproduct.
In ref.\cite{10}
the $\beta(P_0)$ function was chosen to be
\[
\beta(P_0)=M\sin(\frac{P_0}{M})
\]
and the $P_0$ modulo periodicity was restricted to be in the interval
$(-\frac{\pi M}{2},\frac{\pi M}{2})$. This was a first attempt to improve
the ultraviolet behaviour of theories defined on the dPA  introducing
an upper cut-off in the energy spectrum by choosing a bounded function
$\beta(P_0)$. However, when considering the
additivity properties of energy $P_0^{(12)}$ of a system $S^{(12)}$ composed
 of two non-interacting systems $S^{(1)},S^{(2)}$ some problems were found.
Specifically, although the energy is conserved the energy $P^{(12)}$ was no
longer the sum of the energies $P_0^{(1)},P_0^{(2)}$ of the two subsystems
$S^{(1)},S^{(2)}$, respectively. So it was conjectured that the law of
addition of the energies should be
\[
\sin(\frac{P_0^{(1)}}{M})+\sin(\frac{P_0^{(2)}}{M})=2\sin(\frac{P_0^{(12)}}{2M})
.\]
The above conjecture, however,  does not correspond to
 a true coproduct of the generator
 $P_0$ \footnote {We would like to thank
 L. Alvarez-Gaum\'e and O. Ogievetsky for pointing this to us}.
 On the other hand,
the above choice of $\beta(P_0)$  has also a positive aspect in the sense
 that the
 transformation $P_0\rightarrow \beta(P_0)$ is not invertible since
$\beta(P_0)$ is a multivalued function of $P_0$.

Here we would like to discuss two more alternatives as examples of the variety
of existing possibilities.
Let us first assume that the function $\beta(P_0)$  is of the form
\begin{equation}
\beta(P_0)=M\tanh^{-1}(\frac{P_0}{M}). \label{tanh}
\end{equation}
In this case the coproduct of the generators of the dPA are found to be
\begin{eqnarray}
\Delta(J_i)&=&J_i\otimes 1+1\otimes J_i, \nonumber \\
\Delta(K_i)&=&K_i\otimes 1+1\otimes K_i, \nonumber \\
\Delta(P_i)&=&P_i\otimes 1+1\otimes P_i,\nonumber \\
\Delta(P_0)&=&(P_0\otimes 1+1\otimes P_0)(1\otimes 1+\frac{P_0}{M}
\otimes \frac{P_0}{M})^{-1}.
\label{co}
\end{eqnarray}

The coproduct (\ref{co})
is determined by using the property that the transformation
$P_0\rightarrow \beta(P_0)$ transforms the dPA to the ordinary PA,
which in the present case is invertible. Then the knowledge of the coproduct
$\Delta(P_0)$ within PA easily gives us the $\Delta(P_0)$ for the dPA.
At this point it should be emphasized that with the present choice of
$\beta(P_0)$ the dPA is not a trivial redefinition of the ordinary PA.
The reason is that the function $\alpha(P_0)$ is given by
\begin{eqnarray}
\alpha(P_0)=\frac{1}{\beta'(P_0)}=1-\frac{P_0^2}{M^2} .
\end{eqnarray}
It is then clear that states with energy $P_0^2\geq M^2$
 are not representations
of the dPA since the action of the boosts $K_i$  on them
would either leave unchanged
or would reduce their energy. Therefore there is no one-to-one
correspondence among the representations of dPA and PA as  would be in the
case of a trivial redefinition.

As a second example let us assume that the function $\beta(P_0)$ is
\begin{equation}
\beta(P_0)=M\tan^{-1}(\frac{P_0}{M}) .  \label{tan}
\end{equation}
In this case again the coproduct
is determined as before using the
property that the transformation $P_0\rightarrow \beta(P_0)$
takes the dPA to ordinary PA which is
again invertible.  The $\Delta(P_0)$ now  becomes
\begin{equation}
\Delta(P_0)=(P_0\otimes 1+1\otimes P_0)(1\otimes 1-\frac{P_0}{M}
\otimes \frac{P_0}{M})^{-1}.
\end{equation}
 Note however that
the function $\alpha(P_0)$
now is
\begin{equation}
\alpha(P_0)=1+\frac{P_0^2}{M^2}
\end{equation}
and thus it
 does not put any restrictions on the $P_0$'s.
 Therefore there exists  a one-to-one
correspondence among the dPA and PA.

\section{Scalar field theory on dPA}

Here we shall examine a simple field theory such as $\lambda \phi^4$ on the
dPA.
The two examples for construction of multiparticle states  discussed above
will be
examined separately.
Let us consider the ordinary $\phi^4$ in ordinary PA \cite{15}.
 The Lagrangian is
\[
L=L_0+L_I
\] with
\[
L_0= \frac{1}{2}(\partial \phi_0)^2 -\frac{\mu_0^2}{2}\phi_0^2
\]
and
\[
L_I=-\frac{\lambda_0}{4!}\phi_0^4.
\]
Recall that the self-energy graph at 1-loop is given by
\begin{eqnarray}
-\Sigma(p^2)=-\frac{i\lambda_0}{2}\int d^4\ell
\frac{i}{\ell_0^2-\vec{\ell}^2-\mu_0^2+i\epsilon}
\end{eqnarray}
 and it is quadratically divergent. Also the vertex corrections at 1-loop
are  given by
\begin{eqnarray}
\Gamma(s)=
(-\frac{i\lambda_0}{2})^2\int \frac{d^4\ell}{(2\pi)^4}
\frac{i}{(\ell_0-p_0)^2-(\vec{\ell}-\vec{p})^2-\mu_0^2+i\epsilon}
\frac{i}{\ell_0^2-\vec{\ell}^2-\mu_0^2+i\epsilon}
\end{eqnarray}
and  $\Gamma(t),\Gamma(u)$ have similar expressions where
\[
s=p^2=(p_1+p_2)^2\, , \, t=(p_1-p_3)^2\, , \, u=(p_1-p_4)^2
\]
are the Mandelstam variables. The vertex corrections diverge logarithmically.
The divergences of the self-energy and vertex corrections are
 removed by the well-known procedure of introducing
counterterms and performing the renormalization
program. We would like  to examine whether  the same theory when defined
on the dPA has a better ultraviolet behaviour.

Given the form of $\beta(p_0)$ the $L_0$ part of the Lagrangian  can
be expressed
locally in  momentum space as
\begin{equation}
L_0=\frac{1}{2}\left( \tilde\phi_0(p)(\beta^2(p_0)-\vec{p}^2)
\tilde\phi_0(p)-
\mu_0^2\tilde\phi_0(p)\tilde\phi_0(p)\right),
\end{equation}
where $\tilde\phi_0(p)$ is the Fourier transform of $\phi_0(x)$.
Therefore the propagator in  the dPA is
\[
\frac{i}{\beta^2(p_0)-\vec{p}^2-\mu_0^2+i\epsilon},
\]
which with an  appropriate choice of $\beta(p_0)$ can have  more convergent
 behaviour for large $p_0$ as compared to the usual
one
\[
\frac{i}{p_0^2-\vec{p}^2-\mu_0^2+i\epsilon}.
\]

The calculation of 1-loop graphs reduces in determining integrals of the form
\begin{equation}
I(p^2)=\int d^4\ell \beta^{\prime}(\ell_0)
f(\beta^2(\ell_0)-\vec{\ell}^2
,\beta^2(p_0)-\vec{p}^2)
\end{equation}
i.e., integrals with dPA-invariant measure.
Let us start with the case that $\beta(P_0)$ has the form (\ref{tanh}).
In this case the integrals resulting from 1-loop corrections will be
\begin{eqnarray}
I_1=\int_{-M}^{M} d\ell_0 \beta^{\prime}(\ell_0)
\int_{-\infty}^{\infty} d^3\vec{\ell}f(\beta^2(\ell_0)-\vec{\ell}^2)\nonumber
 \\
=\int_{-\infty}^{\infty} d\beta
\int_{-\infty}^{\infty} d^3\vec{\ell}f(\beta^2-\vec{\ell}^2)
\end{eqnarray}
i.e., they are  exactly the same  as in  the ordinary PA case. Therefore
the present form of $\beta(P_0)$ and the corresponding
 dPA invariant measure does not improve the ultraviolet properties
 of the 1-loop
corrections to the theory.

Let us then turn to our second example which exhibits a
different behaviour. In this case $\beta(P_0)$ is given by
eq(\ref{tan}). Then the integrals involved
in the calculations of 1-loop corrections are
\begin{eqnarray}
I_2=\int_{-\infty}^{\infty} d\ell_0 \beta^{\prime}(\ell_0)
\int_{-\infty}^{\infty} d^3\vec{\ell}f(\beta^2(\ell_0)-\vec{\ell}^2) \\
=\int_{-M}^{M} d\beta
\int_{-\infty}^{\infty} d^3\vec{\ell}f(\beta^2-\vec{\ell}^2) ,
\end{eqnarray}
which are clearly more convergent (in view of the cut-off M) than the
corresponding ones in the ordinary PA case.
Therefore this choice of $\beta(P_0)$ provides us with an example of
how one can improve the ultraviolet behaviour of a theory.
 The negative aspect of this particular choice of $\beta(P_0)$
 is that the corresponding dPA has representations in
one-to-one correspondence with  ordinary PA
and one would
 obtain the same results just by
 changing $P_0$ to $\beta(P_0)$ in ordinary PA.
Therefore  this last example  cannot be considered  seriously
as having physical consequences
but rather should be viewed as a regulator of the theory.

\section{A satisfactory model}

Here we present a choice of the function   $\beta(P_0)$
which  seems very promising since
 it satisfies the two new constraints we have
demanded so far. Namely there exists a  coproduct and the
 representations  of the corresponding dPA are different from those of the
ordinary PA. So we can construct a multiparticle state and the
dPA under consideration is a distinct entity separate from the ordinary PA.
Then in principle
the theory defined on this
dPA  could have different physical implications as compared to the same theory
defined on ordinary PA.
As we shall see,
 the one-loop self-energy and vertex corrections of a scalar
field theory defined on this dPA  has improved ultraviolet properties.

The chosen function is
\begin{equation}
\beta(P_0)=M\sin^{-1}(\frac{P_0}{M}). \label{sin}
\end{equation}
 The coproduct of the generators of the dPA
are found to be
\begin{eqnarray}
\Delta(J_i)&=&J_i\otimes 1+1\otimes J_i \nonumber \\
\Delta(K_i)&=&K_i\otimes 1+1\otimes K_i \nonumber \\
\Delta(P_i)&=&P_i\otimes 1+1\otimes P_i \nonumber \\
\Delta(P_0)&=&P_0\otimes \sqrt{1-\frac{P_0^2}{M^2}}
+\sqrt{1-\frac{P_0^2}{M^2}}\otimes P_0.
\end{eqnarray}

As far as the representations  are concerned, recall that the closure of
the algebra requires  that eq.(\ref{1}) should  hold
which in turn implies that
\begin{equation}
\alpha(P_0)=\sqrt{1-\frac{P_0^2}{M^2}}.
\end{equation}
It is then  clear  that the representations with $P_0^2\geq M^2$
 are necessarily non-unitary while   the energy spectrum
for the unitary representations of the dPA
 lies in the interval $(-M,M)$. Therefore, since there is no unitary
representations describing physical states of the dPA with $P_0^2\geq M^2$,
there is
no one-to-one correspondence with the PA.

Turning  to the one-loop self-energy and vertex
 corrections of the scalar theory we find that the self-energy graph
 becomes now
\begin{eqnarray}
-\Sigma(p^2)=-\frac{i\lambda_0}{2}\int_{-\infty}^{\infty} d^4\ell
 \beta^{\prime}(\ell_0)\frac
{i}{\beta^2(\ell_0)-\vec{\ell}^2-\mu_0^2+i\epsilon}\nonumber  \\
=-\frac{i\lambda_0}{2}\int_{-M\pi/2}^{M\pi/2} d\beta \int_{-\infty}^{\infty}
d^3\vec{\ell}
\frac{i}{\beta^2-\vec{\ell}^2-\mu_0^2+i\epsilon}
\end{eqnarray}
which is linearly divergent instead of quadratically in  the usual scalar
 theory defined on
 the ordinary PA. Correspondingly, the one-loop vertex corrections
take the form
\begin{eqnarray}
\Gamma(s)=
(-\frac{i\lambda_0}{2})^2\int \frac{d^4\ell}{(2\pi)^4}
 \beta^{\prime}(\ell_0)
\frac{i}{(\beta(\ell_0)-\beta(p_0))^2-(\vec{\ell}-\vec{p})^2-\mu_0^2+i\epsilon}
\nonumber \\
\frac{i}{\beta^2(\ell_0)-\vec{\ell}^2-\mu_0^2+i\epsilon}\nonumber  \\
=(-\frac{i\lambda_0}{2})^2\int_{-M\pi/2}^{M\pi/2}
 \frac{d\beta d^3\vec{\ell}}{(2\pi)^4}
 \frac
{i}{(\beta(\ell_0)-\beta(p_0))^2-(\vec{\ell}-\vec{p})^2-\mu_0^2+i\epsilon}
\nonumber \\
\frac{i}{\beta^2(\ell_0)-\vec{\ell}^2-\mu_0^2+i\epsilon} .
\end{eqnarray}
and similar forms take the  $\Gamma(t),\Gamma(u)$ which  are all convergent.

\section{Discussion}

The aim of the present paper was to present deformations of the Poincar\'e
algebra that preserve the Lorentz sub-algebra requiring additional constraints
that would pave the way for constructing realistic theories with
improved ultraviolet properties. The new constraints that we have imposed
are (i) the requirement of  the
existence of  a coproduct of the representations
of the dPA and (ii) the demand that there is no one-to-one correspondence
among the representations of the dPA and ordinary PA which means that the
two algebras are just homomorphic.

 Concerning the first requirement,
one may state, as a general rule, that a coproduct for the
dPA always exists if $\beta(P_0)$ is an unbounded function of $P_0$.
In that case, one may employ the homomorphism between the dPA and PA
to pull back the coproduct of the  PA into the
  dPA. Furthermore, if $\beta(P_0)$
is an odd function of $P_0$, the same homomorphism can also pull back
the antipode of the  PA into the  dPA turning the latter into a cocommutative
Hopf algebra.
{}From the physical point of view, the cocommutativity of the dPA guarantees
that the addition of observables for two systems $S^{(1)}$ and $S^{(2)}$ is
independent of the order of addition.
Recall for comparison that in the case of  non-cocommutative
algebras (quantum groups),
 the addition depends on the order (i.e., on the ``labeling").
 Turning to the second requirement it guarantees that the dPA is not a
simple redefinition of ordinary PA.

Although the additional
constraints  are necessary in order to construct a field theory
on a dPA which is not a trivial redefinition of the ordinary PA  they
do not guarantee
that the theory has better ultraviolet behaviour. Therefore from
this point of view they are
necessary but not sufficient. On the other hand all the constraints
considered so far, including the requirement  for improved ultraviolet
behaviour, cannot restrict in an appreciable manner the choices of the
functions $\beta(P_0)$ which differentiate the various dPAs from
each other.

We should emphasize that when a dPA satisfying
 all the above constraints is found
it has very important physical consequences. First of all,  such a dPA will
be characterized by a non-trivial function $\beta(P_0)$ which certainly will
result in observable
 deviations of the special theory of relativity.
 There exist already some analyses \cite{16,17}
which put limits on the characteristic mass scale M
appearing in general in $\beta(P_0)$ on dimensional grounds. For instance,
according to ref.\cite{17}
the lowest bound consistent with experimental observations is $M_{min}\simeq
10^{12}$GeV.

Finally, a field theory with less divergences than the usual ones requires
also  less counterterms to cure them. This in turn means
that the theory will have less free parameters to be
 fixed by experiment, or equivalently the theory
will have more predictive power.
It is expected  then that the phenomenological constraints will provide us
with enough information to restrict the possible choices of the functions
$\beta(P_0)$. Moreover, it is fair to hope that genuine predictions on
unkown parameters would emerge as a result of the above construction.

\vspace{1cm}

{\bf Acknowledgement} \\

We would like to thank M. Arik, E. Kiritsis,  J. Kubo,
M. Niedermaier, O. Ogievetsky
and S. Theisen for useful discussions.

\end{document}